\documentclass[a4paper,12pt]{article}

\usepackage{bbold}
\usepackage{float}
\usepackage{cancel}
\usepackage{subfig}
\usepackage{amsmath, amssymb, setspace,cite}
\usepackage{graphicx}
\usepackage{hyperref}
\usepackage{color}
\usepackage{soul}
\usepackage{xcolor}
\usepackage{multirow}

\vspace{0.6cm}
\date{\today} 

\usepackage{tabularx}
\usepackage{longtable} 
\usepackage{multicol}
\usepackage{multirow}

\renewcommand{\arraystretch}{1.2}

\definecolor{dgreen}{rgb}{0.0, 0.5, 0.0}

\begin{document}


\def\thefootnote{\fnsymbol{footnote}}

\begin{center}
\Large\bf\boldmath
Improving lepton flavour universality tests\\ with $K_L$ decays
\end{center}
\vspace{0.6cm}

\begin{center}
G.~D'Ambrosio$^{1}$\footnote{Electronic address: gdambros@na.infn.it}, 
A.M. ~Iyer$^{2}$\footnote{Electronic address: iyerabhishek@physics.iitd.ac.in}, F.~Mahmoudi$^{3,4,5}$\footnote{Electronic address: nazila@cern.ch}, 
S. Neshatpour$^{3}$\footnote{Electronic address: s.neshatpour@ip2i.in2p3.fr}\\
\vspace{0.6cm}
{\sl $^1$INFN-Sezione di Napoli, Complesso Universitario di Monte S. Angelo,\\ Via Cintia Edificio 6, 80126 Napoli, Italy}\\[0.4cm]
{\sl $^2$Department of Physics, Indian Institute of Technology Delhi,\\ Hauz Khas, New Delhi-110016, India}\\[0.4cm]
{\sl $^3$Université Lyon 1, CNRS, IP2I, UMR 5822, Villeurbanne, France}\\[0.4cm]
{\sl $^4$Theoretical Physics Department, CERN, CH-1211 Geneva 23, Switzerland}\\[0.4cm]
{\sl $^5$Institut Universitaire de France (IUF), 75005 Paris, France }\\[0.4cm]
\end{center}
\renewcommand{\thefootnote}{\arabic{footnote}}
\setcounter{footnote}{0}

\begin{abstract}

Rare kaon decays provide sensitive probes of the flavour structure of the Standard Model and of possible new physics. We perform a global analysis incorporating recent experimental results and updated Standard Model predictions, including the latest measurement of $K^+ \to \pi^+ \nu\bar{\nu}$ and lepton flavour universality observables in $K^+ \to \pi^+ \ell^+\ell^-$. The fit favours a best-fit point close to the Standard Model, while a second local minimum remains phenomenologically relevant. We define benchmark scenarios associated with these two regions and investigate the prospective sensitivity of NA62 and KOTO-II to the new physics parameter space. We consider projected measurements of $K^+ \to \pi^+ \nu\bar{\nu}$ and lepton flavour universality observables in $K^+ \to \pi^+ \ell^+\ell^-$ at NA62, and of $K_L \to \pi^0 \nu\bar{\nu}$, $K_L \to \pi^0 e^+e^-$, and $K_L \to \pi^0 \mu^+\mu^-$ at KOTO-II. We find that KOTO-II has significant potential to probe and discriminate between the viable new physics scenarios, with NA62 providing complementary sensitivity.

\end{abstract}

\clearpage 
\section{Introduction}

Rare kaon decays provide a particularly powerful framework for testing the flavour structure of the Standard Model (SM) and for probing possible contributions from physics beyond it. At present, NA62 with its programmes on $K^+\rightarrow\pi^+\nu \bar\nu$ and $K^+\rightarrow\pi^+\ell^+\ell^-$, LHCb with its search for $K_S\rightarrow\mu^+\mu^-$, and KOTO targeting $K_L\rightarrow\pi^0\nu \bar\nu$, are at the forefront of experimental efforts in kaon physics. 
Among them, the $K \to \pi \nu \bar{\nu}$ decays play a distinguished role. Their amplitudes are dominated by short-distance contributions and can be predicted with high theoretical precision, making them particularly clean probes of the flavour sector. Both the charged and neutral modes, $K^+ \to \pi^+ \nu \bar{\nu}$ and $K_L \to \pi^0 \nu \bar{\nu}$, respectively, are sensitive to complementary combinations of the underlying short-distance couplings that lead to a high degree of correlation between their respective theoretical predictions \cite{Grossman:1997sk}.

These experiments have obtained remarkable results. For instance,
the NA62 experiment has recently reported an updated measurement of the branching ratio of $K^+ \to \pi^+ \nu \bar{\nu}$ \cite{NA62_LaThuile_2026}. In parallel, the KOTO experiment continues to improve the sensitivity to the decay $K_L \to \pi^0 \nu \bar{\nu}$~\cite{KOTO:2024zbl}, while the proposed KOTO-II experiment aims to substantially extend the experimental reach for this and other rare neutral-kaon decay modes~\cite{KOTO:2025gvq}. The renewed focus on the electron channel in $K^+\rightarrow\pi^+\ell^+\ell^-$, permits the reinvestigation of  
lepton flavour universality violating (LFUV) observable $(a_+^{\mu\mu}-a_+^{ee})$, where $a_+^{\ell\ell}$ is the leading order term in the parametrisation of the vector form factor \cite{Crivellin:2016vjc}.  In comparison with earlier analyses, recent studies suggest an increasing compatibility with the SM predictions~\cite{DAmbrosio:2018ytt,NA62:2022qes}.

These measurements also quantify the extent of possible contributions of Beyond the Standard Model (BSM) operators.
Using the framework developed for studying the  global impact of kaon decays on the Wilson coefficients of the effective operators~\cite{DAmbrosio:2022kvb,DAmbrosio:2023irq,DAmbrosio:2024ewg},
we study the parameter space of (SM and BSM) Wilson coefficients and the impact of future measurements.

The present analysis incorporates the latest available experimental information together with updated SM predictions based on the most recent determinations of the relevant quark masses and Cabibbo--Kobayashi--Maskawa (CKM) parameters~\cite{ParticleDataGroup:2026aaa}. It identifies a degeneracy in the parameter space of the Wilson coefficients that would not be resolvable by NA62. This paves the way for future exploration in kaon physics and particularly KOTO-II, where investigations of the decay of $K_L$ into both neutrinos and charged leptons can break this degeneracy.

The remainder of this paper is organised as follows. In Section~\ref{sec:Framework} the framework as well as the current situation is discussed. In Section~\ref{sec:projections} the projections for possible future measurements are presented. We conclude in Section~\ref{sec:conclusions}.   

\section{Framework}
\label{sec:Framework}

\subsection{Effective Hamiltonian}
\label{subsec:EffHam}
The starting point is the following effective hamiltonian that parametrises the $s\to d$ transitions~as:

\begin{equation}
\mathcal{H}_{\rm eff}
=
-\frac{4G_F}{\sqrt{2}}
\lambda_t^{sd}
\frac{\alpha_e}{4\pi}
\sum_k
C_k^\ell
O_k^\ell \, ,
\label{eq:Heff}
\end{equation}
where $\lambda_t^{sd}=V_{ts}^*V_{td}$. In the most general effective field theory, the sum extends over all operators compatible with the symmetries of the low-energy theory. In this work, however, we restrict our analysis to the operators:
\begin{align}
O_9^\ell
&=
(\bar s\gamma_\mu P_L d)
(\bar\ell\gamma^\mu\ell)\, ,
\nonumber\\
O_{10}^\ell
&=
(\bar s\gamma_\mu P_L d)
(\bar\ell\gamma^\mu\gamma_5\ell)\, ,
\nonumber\\
O_L^\ell
&=
(\bar s\gamma_\mu P_L d)
(\bar\nu_\ell\gamma^\mu(1-\gamma_5)\nu_\ell)\, ,
\label{eq:operators}
\end{align}
with $P_L=(1-\gamma_5)/2$. The corresponding Wilson coefficients are written as:
\begin{equation}
C_k^\ell
= C_{k,\rm SM}^\ell + \delta C_k^\ell\, ,
\end{equation}
where $C_{k,{\rm SM}}^\ell$ and $\delta C_k^\ell$ parametrise the SM and possible new physics contributions, respectively.

Although the complete operator basis also contains the chirality-flipped version of the above operators, as well as scalar and pseudoscalar operators~\cite{DAmbrosio:2025amb}, these are not considered in the present analysis. Instead, we focus on the left-handed semileptonic operators, which are directly connected to the framework commonly used in studies of flavour anomalies in the $B$ sector. Following Ref.~\cite{DAmbrosio:2022kvb} we assume that the charged and neutral leptons belong to the same $SU(2)_L$ doublets. The Wilson coefficients thus satisfy:
\begin{equation}
\delta C_L^\ell
= \delta C_9^\ell
= -\delta C_{10}^\ell \, .
\end{equation}
Consequently, all results are presented in terms of the coefficients $\delta C_L^\ell$.
As this work  also investigates the sensitivity of future rare kaon measurements to lepton flavour universality violation, we  adopt the following convention:
\begin{equation}
\delta C_L^\mu=\delta C_L^\tau \, ,
\qquad
\delta C_L^e\neq\delta C_L^\mu \, .
\end{equation}
This assumption reduces the fit to two independent Wilson coefficients while retaining the possibility of lepton flavour universality violation.

\subsection{ Methodology and observables}
\label{subsec:FitMethod}

The fit is performed by minimising the $\chi^2$ function:
\begin{equation}
 \chi^2=
 \sum_{i,j}
 \left( O_i^{\rm th}-O_i^{\rm exp} \right)
 C^{-1}_{ij} \left( O_j^{\rm th} - O_j^{\rm exp} \right),
\end{equation}
where $O_i^{\rm th}$ and $O_i^{\rm exp}$ denote the theoretical predictions and experimental measurements, respectively, and $C_{ij}$ is the total covariance matrix including both theoretical and experimental uncertainties. The observables used in the fit are listed in Table~\ref{tab:data}, which includes the current experimental status and the precision for potential future measurements.  The updated Standard Model predictions are computed using \texttt{SuperIso} package~\cite{Mahmoudi:2007vz,Mahmoudi:2008tp,Mahmoudi:2009zz,Neshatpour:2022fak} with latest determinations of the CKM parameters and quark masses~\cite{ParticleDataGroup:2026aaa}.

The interpretation of some of the observables in Table~\ref{tab:data} depends on the treatment of long-distance contributions. In the case of $K_L\to\mu^+\mu^-$, the decay amplitude receives both a short-distance contribution and a long-distance contribution arising from the two-photon intermediate state, $K_L\to\gamma\gamma\to\mu^+\mu^-$.
The sign of the amplitude for $K_L\to\gamma\gamma$ is not experimentally determined, resulting in an ambiguity in the interference between the short- and long-distance contributions. A positive (negative) sign corresponds to destructive (constructive) interference in the branching ratio. In this work we adopt the positive solution, denoted by $(+)$ in Table~\ref{tab:data}, which yields a Standard Model prediction in better agreement with the experimental measurement (see Refs.~\cite{DAmbrosio:1996kjn,DAmbrosio:1997eof,Isidori:2003ts,Gerard:2005yk,Hoferichter:2023wiy,DAmbrosio:2017klp} for discussions of the sign ambiguity).

Another sign ambiguity is present in the decays $K_L\to\pi^0\ell^+\ell^-$. Their branching ratios receive contributions from direct CP violation, indirect CP violation through $K^0$--$\bar K^0$ mixing, and a CP-conserving long-distance contribution originating from two-photon exchange. In addition, the direct and indirect CP-violating amplitudes interfere, with either constructive or destructive interference being possible. Following the current theoretical preference~\cite{Bruno:1992za,Buchalla:2003sj}, we assume constructive interference throughout this work, corresponding to the $(+)$ entries in Table~\ref{tab:data}.

\begin{table}[!t]
\renewcommand{\arraystretch}{1.39}
\begin{center}
\setlength\extrarowheight{1pt}
\scalebox{0.73}{
\hspace*{-2mm}
\begin{tabular}{|llll|c
|}\hline\hline
\bf{Observable} & \bf{SM prediction}& \bf{Experimental results} & \bf{Reference}&   \textbf{Precision for projections} \\ \hline
BR$(K^+\to \pi^+\nu\bar\nu)$    & $(8.38 \pm 0.60)\times 10^{-11}$  & $(9.6^{+1.8}_{-1.6}\vert_{\rm stat}\;^{+0.8}_{-0.6}\vert_{\rm syst} ) \times 10^{-11}$ & ~\cite{NA62_LaThuile_2026}   & $15\%$ \cite{JParcKaon2024} \\
BR$(K^0_L\to \pi^0\nu\bar\nu)$  & $(3.00 \pm 0.29) \times 10^{-11}$ & $ <2.2\times 10^{-9}$ @$90\%$ CL & ~\cite{KOTO:2024zbl} & $25\%$ \cite{KOTO:2025gvq}\\
LFUV($a_+^{\mu\mu}-a_+^{ee}$)&\multicolumn{1}{c}{0}&$-0.014\pm 0.016$&~\cite{DAmbrosio:2018ytt,NA62:2022qes} & $0.009$  \\
BR$(K_L\to \mu\bar\mu)$ ($+$)   & $(6.84^{+0.77}_{-0.29})\times 10^{-9}$    & \multirow{2}{*}{$(6.84\pm0.11)\times 10^{-9}$} & \multirow{2}{*}{\cite{ParticleDataGroup:2024cfk}} &  \multirow{2}{*}{Current } \\
BR$(K_L\to \mu\bar\mu)$ ($-$)   &  $ (8.11^{+1.47}_{-0.98})\times 10^{-9}$      &  &&
\\
\multirow{2}{*}{BR$(K_S\to \mu\bar\mu)$}         & \multirow{2}{*}{$(5.16\pm1.53)\times 10^{-12}$}    & $ < 2.1(2.4)\times 10^{-10}$ @$90(95)\%$ CL & \multirow{2}{*}{~\cite{LHCb:2020ycd}}  & \multirow{2}{*}{Current

} \\
& & ~~~~$\left( 0.9^{+0.7}_{-0.6}\times 10^{-10} \right)$ &  & \\ 
BR$(K_L\to \pi^0 e\bar e)(+)$         & $(3.56^{+0.92}_{-0.80})\times 10^{-11}$    & \multirow{2}{*}{$ < 28\times 10^{-11}$ @$90\%$ CL} & \multirow{2}{*}{\cite{KTeV:2003sls}} & \multirow{2}{*}{ 25\% \cite{KOTO:2025gvq}}\\
BR$(K_L\to \pi^0 e\bar e)(-)$         & $(1.54^{+0.60}_{-0.48})\times 10^{-11}$        &&& \\
BR$(K_L\to \pi^0 \mu\bar \mu)(+)$         & $(1.41^{+0.27}_{-0.25})\times 10^{-11}$    & \multirow{2}{*}{$ < 38\times 10^{-11}$ @$90\%$ CL} & \multirow{2}{*}{\cite{KTEV:2000ngj}} & \multirow{2}{*}{ 25\% \cite{KOTO:2025gvq}} \\
BR$(K_L\to \pi^0 \mu\bar \mu)(-)$         & $(0.94^{+0.21}_{-0.20})\times 10^{-11}$       &&&  \\
\hline \hline
\end{tabular}}
\caption{\small
The SM predictions, current experimental values, and projected precisions. In the last column, ``Current'' signifies that the measurement precision or the upper bound is maintained at the current experimental level.
\label{tab:data}}
\end{center}
\end{table}

\subsection{Current constraints and global fit}
\label{subsection:GlobalFit}
\begin{figure}[tb!]
\begin{center}
\includegraphics[width=0.7\textwidth]{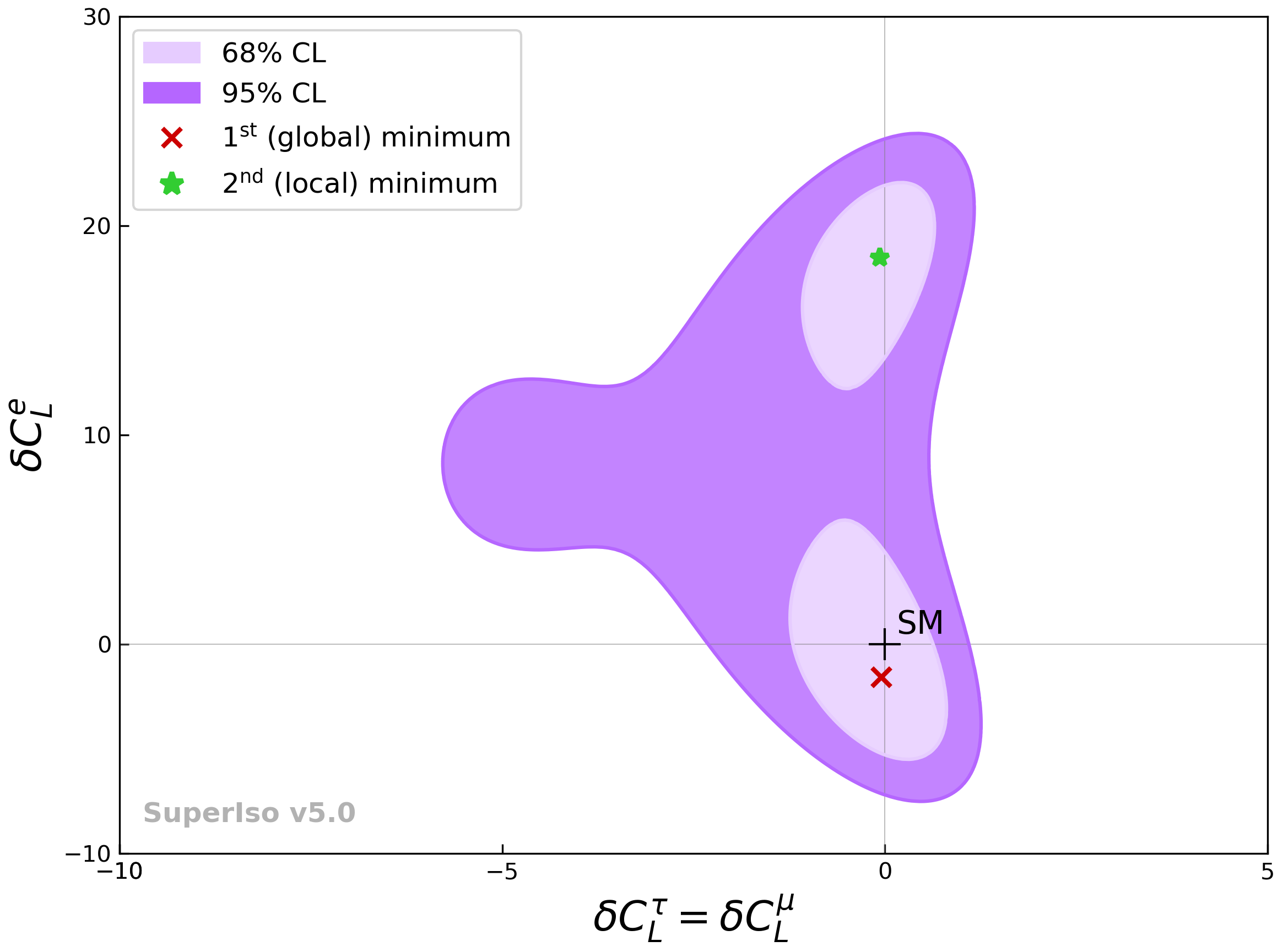}
\vspace{-0.2cm}
\caption{\small Global fit to rare kaon decays using all the data given in Table~\ref{tab:data}, assuming LD:$+$ for BR($K_L\to \mu^+\mu^-$), and constructive interference for BR($K_L \to \pi^0 \ell^+\ell^-$).
\label{fig:CurrentDataFit}}
\end{center}
\end{figure}

The combined impact of the updated SM inputs and the current experimental status of the relevant observables is reflected in the global fit illustrated in Fig.~\ref{fig:CurrentDataFit}.

\begin{figure}[tb!]
\begin{center}
\includegraphics[width=1\textwidth]{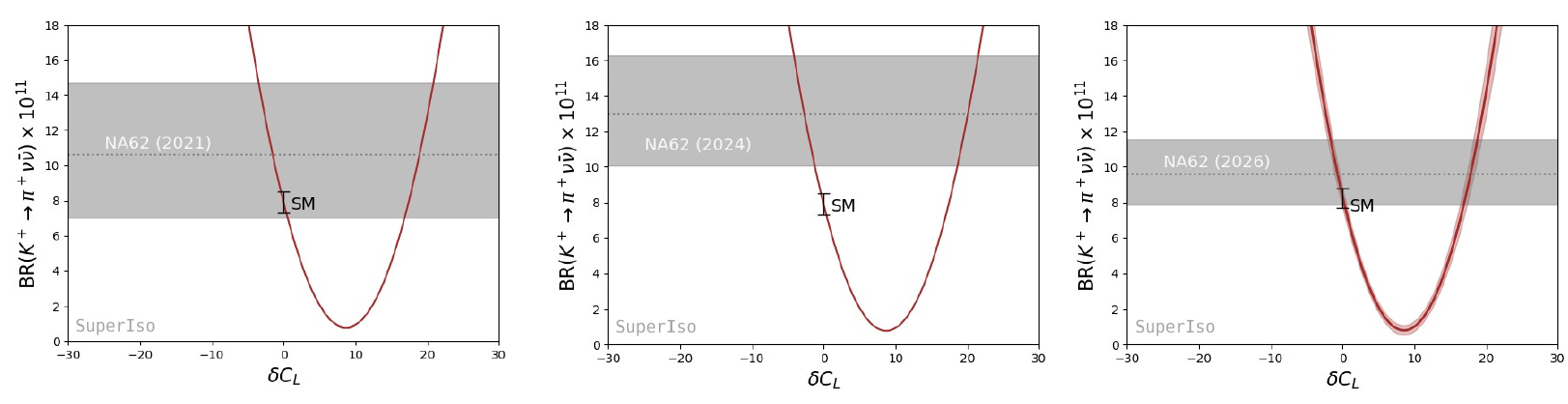}
\vspace{-0.2cm}
\caption{\small Comparison of the impact of new physics contribution to BR$(K^+\rightarrow\pi^+\nu\bar\nu)$ with the evolving NA62 measurement. The experimental bands, left to right, correspond to the NA62 results from 2021~\cite{NA62:2021zjw}, 2024~\cite{NA62:2024pjp}, and 2026~\cite{NA62_LaThuile_2026}, respectively. The theoretical predictions in the left and middle panels are from Ref.~\cite{DAmbrosio:2022kvb}, while the right panel uses the updated prediction from Table~\ref{tab:data} and also includes the $1\sigma$ theoretical uncertainty.
\label{fig:KpinunuCL}}
\end{center}
\end{figure}

Compared to our previous analysis~\cite{DAmbrosio:2022kvb}, we observe consistency of the existing measurements with the SM in the region of the global minimum (lower bubble). 
The global best-fit point lies very close to the Standard Model point, corresponding to $\delta C_L^\ell=0$. 
This is primarily driven by the NA62 updated measurement of $\mathrm{BR}(K^+\to\pi^+\nu\bar{\nu})$~\cite{NA62_LaThuile_2026} and, to a lesser degree, by the NA62 updated measurement~\cite{NA62:2022qes} of the $K^+\to \pi^+\mu^+\mu^-$ decay. This is illustrated in Fig.~\ref{fig:KpinunuCL}, where the plots from left to right show the comparison of the NA62 evolving measurements of BR$(K^+\rightarrow\pi^+\nu\bar\nu)$. The impact of new physics (through flavour universal coupling $\delta C_L$) is also shown, where in the right plot, the theoretical prediction has also changed compared to our previous work~\cite{DAmbrosio:2022jmd} due to updated input parameters, the most important of which are the CKM parameters.
The trajectory of the global best-fit point followed the degree of agreement between the SM prediction and the experimental measurement for BR$(K^+\rightarrow\pi^+\nu\bar\nu)$.

\begin{figure}[htb!]
\begin{center}
\includegraphics[width=0.49\textwidth]{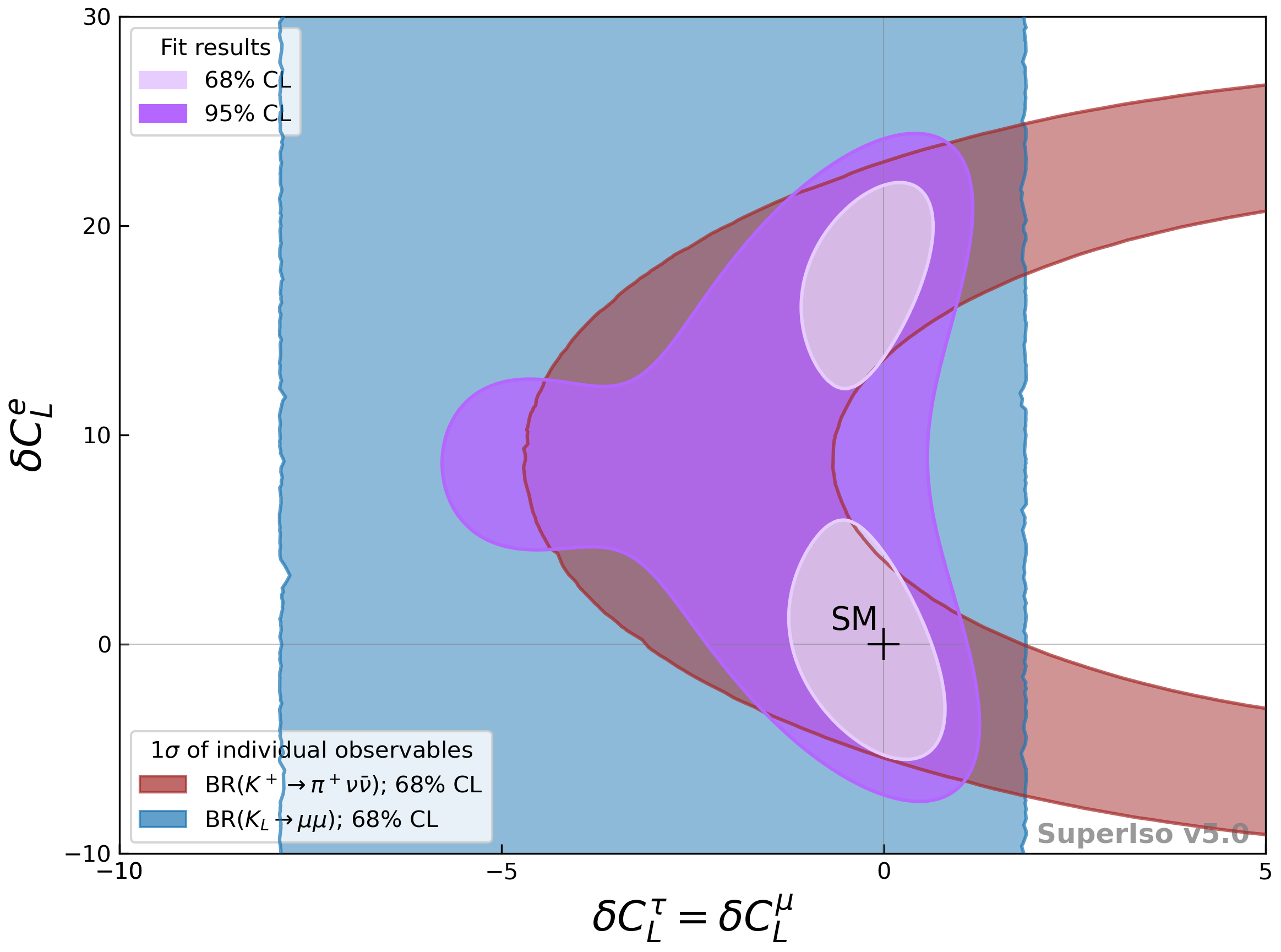}
\includegraphics[width=0.49\textwidth]{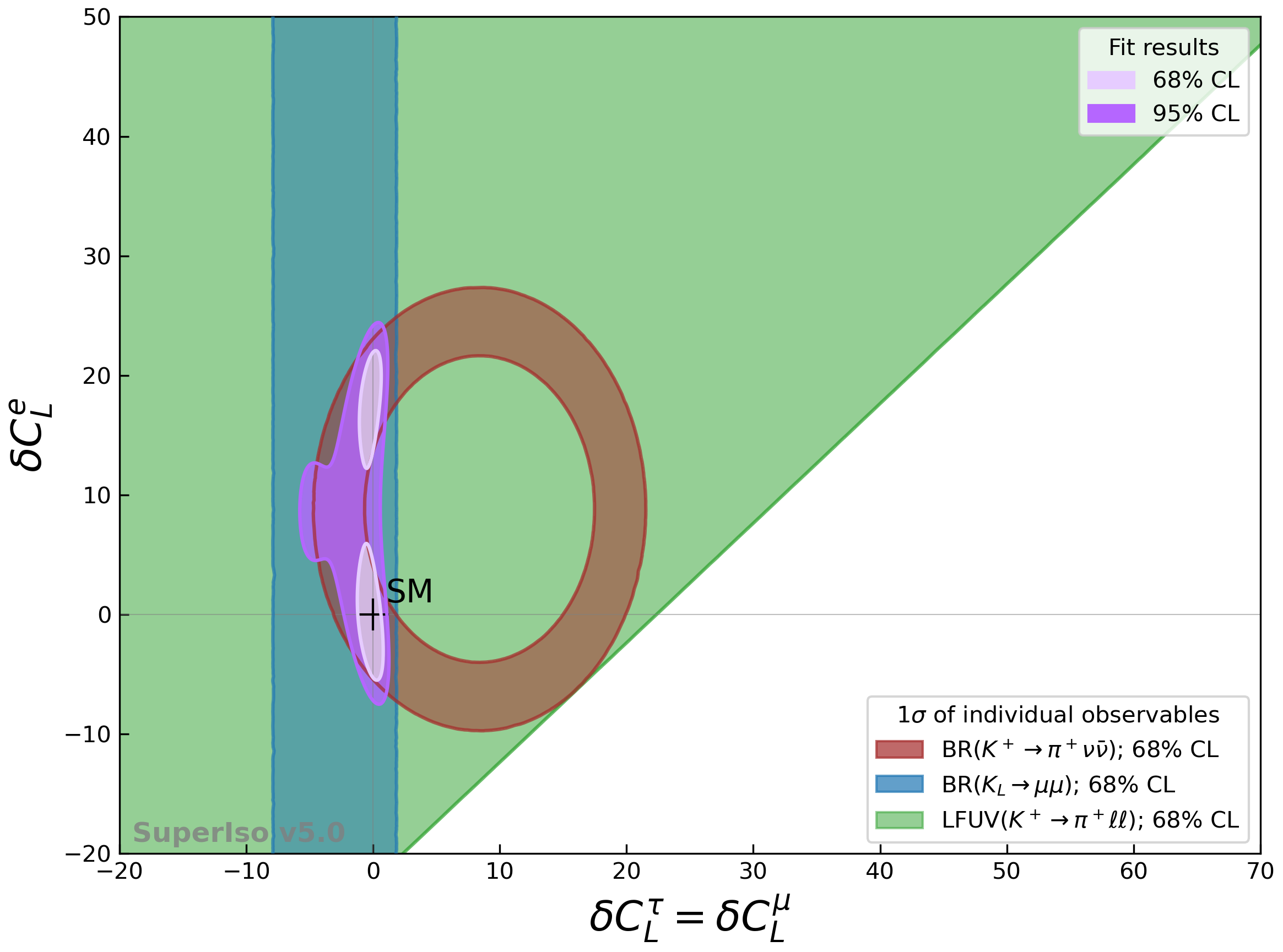}
\vspace{-0.2cm}
\caption{\small Global fit to the current data together with the $68\%$ CL regions obtained from selected individual observables. The right panel shows a wider view of the parameter space, illustrating the full constraint from $\mathrm{BR}(K^+\to\nu\bar\nu)$ and $\mathrm{BR}(K_L\to\mu^+\mu^-)$ as well as the comparatively weaker constraint from the LFUV observable in $K^+\to\pi^+\ell^+\ell^-$.
\label{fig:fit_individual}}
\end{center}
\end{figure}

To illustrate the origin of the allowed regions and the role of the different observables, Fig.~\ref{fig:fit_individual} also shows the constraints obtained from selected individual measurements. As already noted,  the golden channel $K^+\to\pi^+\nu\bar{\nu}$ drives the shape of the parameter space, restricting it to the narrow crescent-shaped region shown in the left panel.


The branching ratio of $K_L\to\mu^+\mu^-$, despite providing a weaker constraint individually, plays an important role in the global fit. This observable probes a different combination of Wilson coefficients and therefore removes the approximate degeneracy between $\delta C_L^e$ and $\delta C_L^\mu=\delta C_L^\tau$ left by $K^+\to\pi^+\nu\bar{\nu}$. This is more clearly visible in the right panel, where the full $68\%$ CL region obtained from $\mathrm{BR}(K_L\to\mu^+\mu^-)$ and $\mathrm{BR}(K^+\to\pi^+\nu\bar\nu)$ is displayed.

The LFUV observable in $K^+\to\pi^+\ell^+\ell^-$ provides a comparatively weaker constraint due to its current experimental precision. Nevertheless, it has an impact on the location of the global best-fit point. In particular, the updated measurement is more compatible with the Standard Model compared to the previous determination~\cite{Bician:2020ukv}. Consequently, the preferred region has moved closer to the Standard Model compared to the result obtained in Ref.~\cite{DAmbrosio:2022kvb}.

The plot in Fig.~\ref{fig:CurrentDataFit} also shows the appearance of another bubble, above the bubble for the global minimum and roughly aligned with the line $\delta C_L^\mu=\delta C_L^\tau=0$. It corresponds to the 68\% CL for the second minimum shown in green. Together, these bubbles show that although both regions are consistent with the existing data, the $68\%$ CL regions have a distinctly visible separation. This hints at the possibility of future statistical discrimination, which we discuss in the next section.

\subsection{Lepton flavour degeneracy of \texorpdfstring{$K\to \pi\nu\bar\nu$}{K->pi nu nu} measurement}

\begin{figure}[t!]
\begin{center}
\includegraphics[width=0.49\textwidth]{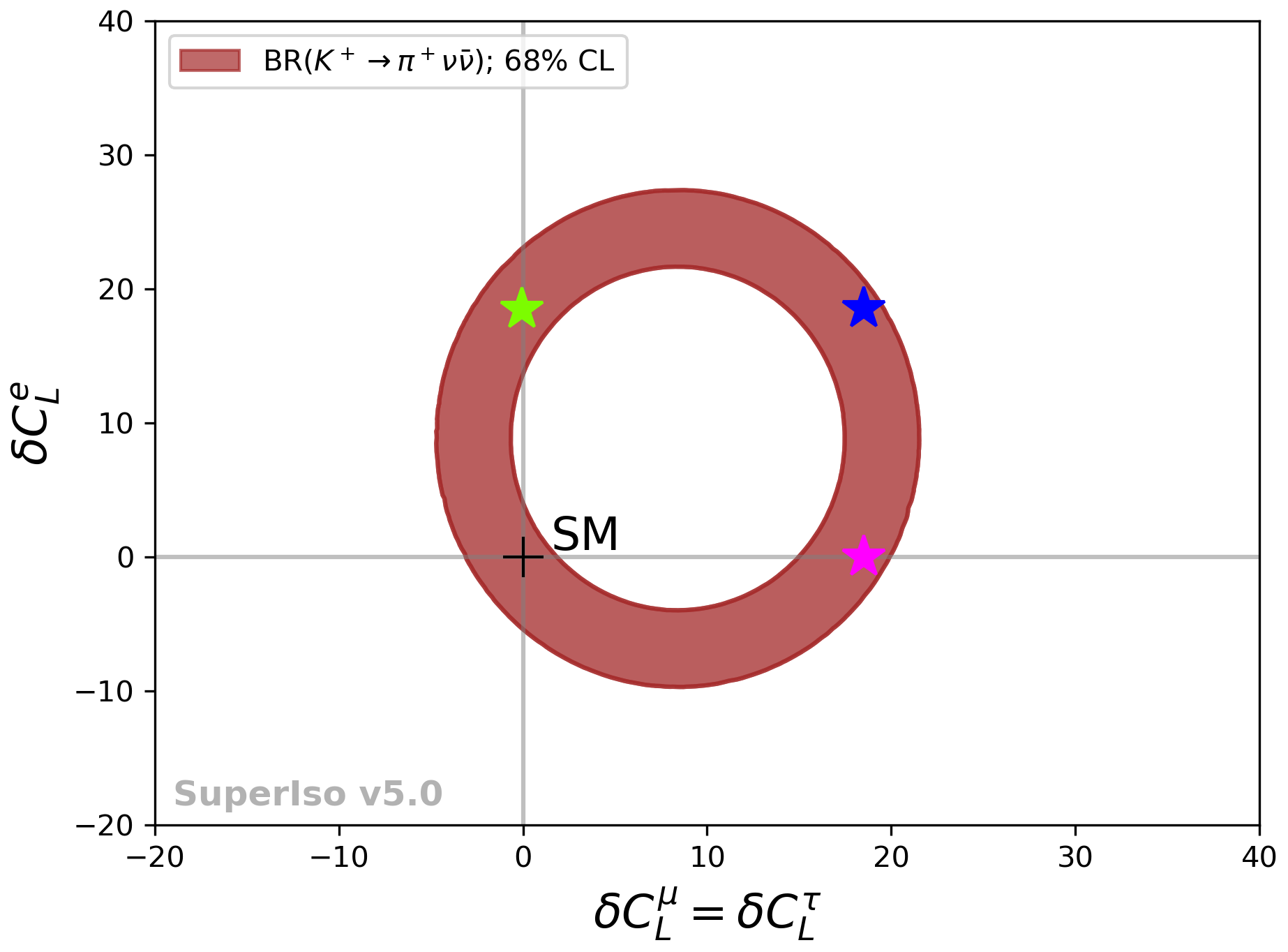}
\vspace{-0.2cm}
\caption{\small
The $68\%$ CL region allowed by $\mathrm{BR}(K^+\to\pi^+\nu\bar\nu)$ in the $(\delta C_L^\mu,\, \delta C_L^e)$ plane. The three stars illustrate representative non-SM points compatible with the NA62 measurement: the blue star is flavour universal, while the green and magenta stars correspond to LFUV scenarios.
\label{fig:Kpinunu_CL}}
\end{center}
\end{figure}

Despite the remarkable agreement of the BR($K^+ \to \pi^+ \nu\bar\nu$) measurement with the SM prediction, LFUV new physics remains compatible with the measurement. This is illustrated in Fig.~\ref{fig:Kpinunu_CL}, where the maroon band shows the region in the $(\delta C_L^\mu,\, \delta C_L^e)$ plane allowed at $68\%$ CL by  BR($K^+\rightarrow\pi^+\nu\bar \nu$). 
The inclusive neutrino final state leaves a degeneracy between the different lepton flavour contributions, allowing sizeable departures from lepton flavour universality while reproducing the observed branching ratio.

The three stars illustrate representative points along this allowed region. The blue star is flavour conserving, whereas the green and magenta stars correspond to LFUV scenarios. In particular, the green point coincides with the second minimum of the global fit shown in Fig.~\ref{fig:CurrentDataFit}. Thus, even a precise determination of $\mathrm{BR}(K^+\to\pi^+\nu\bar\nu)$ alone cannot resolve the different lepton flavour structures compatible with the data. This provides a strong motivation for future measurements, in particular the complementary rare $K_L$ decay modes accessible at KOTO-II.

\section{Breaking the Degeneracy with KOTO-II}
\label{sec:projections}

To assess the ability of future measurements to distinguish the two regions identified in the current global fit as given in Fig.~\ref{fig:CurrentDataFit}, we define two benchmark scenarios corresponding to its global and second local minima. For each benchmark, the central values of the future observables are taken to be those predicted at the corresponding minimum. Specifically, we consider the projected measurements of $\mathrm{BR}(K^+\to\pi^+\nu\bar\nu)$ and the LFUV observables in $K^+\to\pi^+\ell^+\ell^-$ at NA62, as well as $\mathrm{BR}(K_L\to\pi^0\nu\bar\nu)$, $\mathrm{BR}(K_L\to\pi^0\mu^+\mu^-)$, and $\mathrm{BR}(K_L\to\pi^0e^+e^-)$ at KOTO-II. The BR($K_{L} \to \mu^+\mu^-$) and BR($K_{S} \to \mu^+\mu^-$) measurements are kept at their current experimental values.

The two benchmark scenarios are defined as follows:
\begin{itemize}
\item \textbf{Benchmark 1:} the predictions associated with the global minimum of the current fit, denoted by the red cross in Fig.~\ref{fig:CurrentDataFit}.
\item \textbf{Benchmark 2:} the predictions associated with the second local minimum of the current fit, denoted by the green star in Fig.~\ref{fig:CurrentDataFit}.
\end{itemize}

With the current experimental precision, the two benchmarks are consistent at $1\sigma$. Their future discrimination is therefore determined by the extent to which the projected NA62 and KOTO-II measurements can distinguish the predictions of the two solutions. The NA62 measurement of $\mathrm{BR}(K^+\to\pi^+\nu\bar\nu)$ is expected to reach an uncertainty of approximately $15\%$ with the full data set. We also consider improved measurements of the LFUV observables in $K^+\to\pi^+\ell^+\ell^-$. For the latter, we assume uncertainties of $\sigma_{a_+^{\mu\mu}}=\pm0.007$ and $\sigma_{a_+^{ee}}=\pm0.005$ for the muon and electron channels, respectively~\cite{KarimMassri:private}. In addition, the proposed KOTO-II experiment envisages measurements of several rare $K_L$ decays that currently only have upper limits. We assume a projected precision of $25\%$ for each of $\mathrm{BR}(K_L\to\pi^0\nu\bar\nu)$, $\mathrm{BR}(K_L\to\pi^0e^+e^-)$, and $\mathrm{BR}(K_L\to\pi^0\mu^+\mu^-)$~\cite{KOTO:2025gvq}. The projected precisions of all observables are summarised in the last column of Table~\ref{tab:data}.

For each projected measurement, we take the uncertainty specified above and scale it with the projected central value corresponding to the chosen benchmark. We then consider the three experimental programmes separately and in combination:
\begin{itemize}
\item \textbf{NA62 only:} we include the projected NA62 measurements of $\mathrm{BR}(K^+\to\pi^+\nu\bar\nu)$ and the LFUV observables in $K^+\to\pi^+\ell^+\ell^-$, while keeping the KOTO-II observables at their current precision.
\item \textbf{KOTO-II only:} we include the projected KOTO-II measurements of $\mathrm{BR}(K_L\to\pi^0\nu\bar\nu)$, $\mathrm{BR}(K_L\to\pi^0e^+e^-)$, and $\mathrm{BR}(K_L\to\pi^0\mu^+\mu^-)$, while keeping the NA62 observables at their current precision.
\item \textbf{Combined:} we include the projected measurements from both NA62 and KOTO-II.
\end{itemize}

\begin{figure}[t!]
\begin{center}
\includegraphics[width=0.45\textwidth]{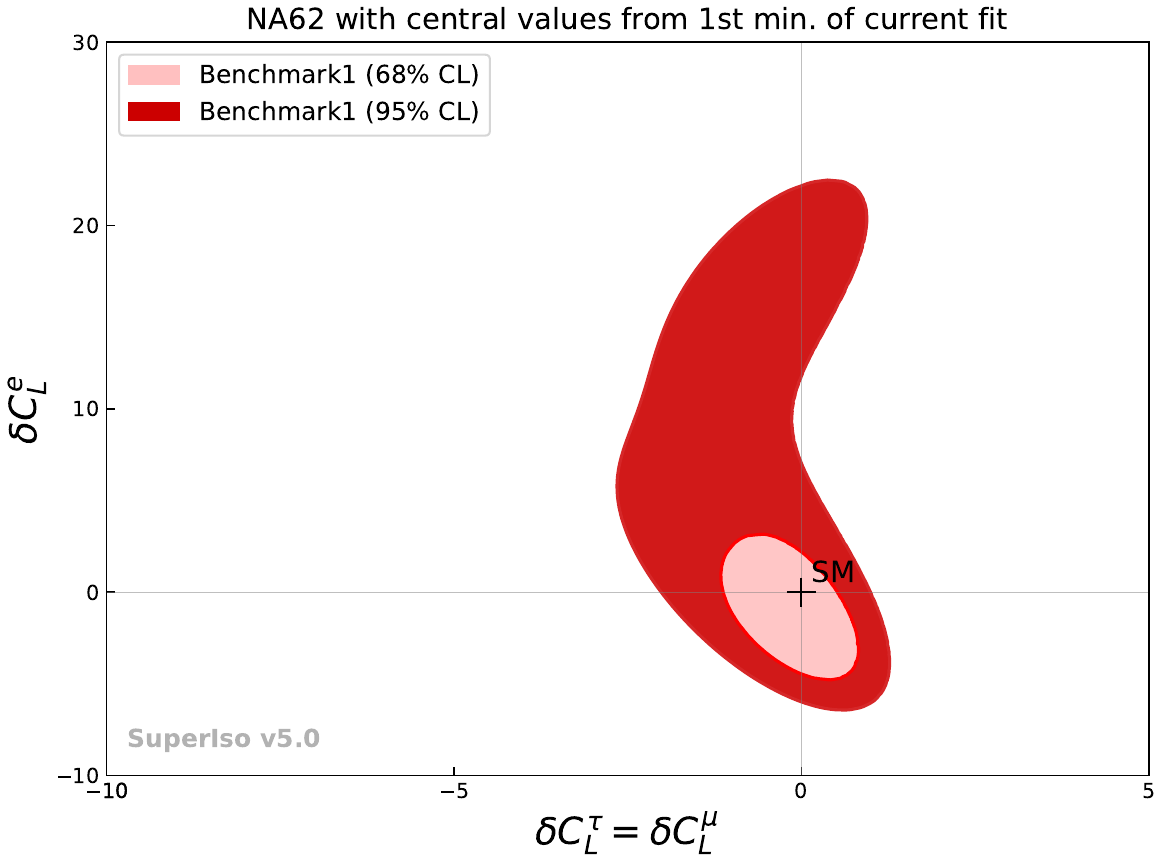}
\includegraphics[width=0.45\textwidth]{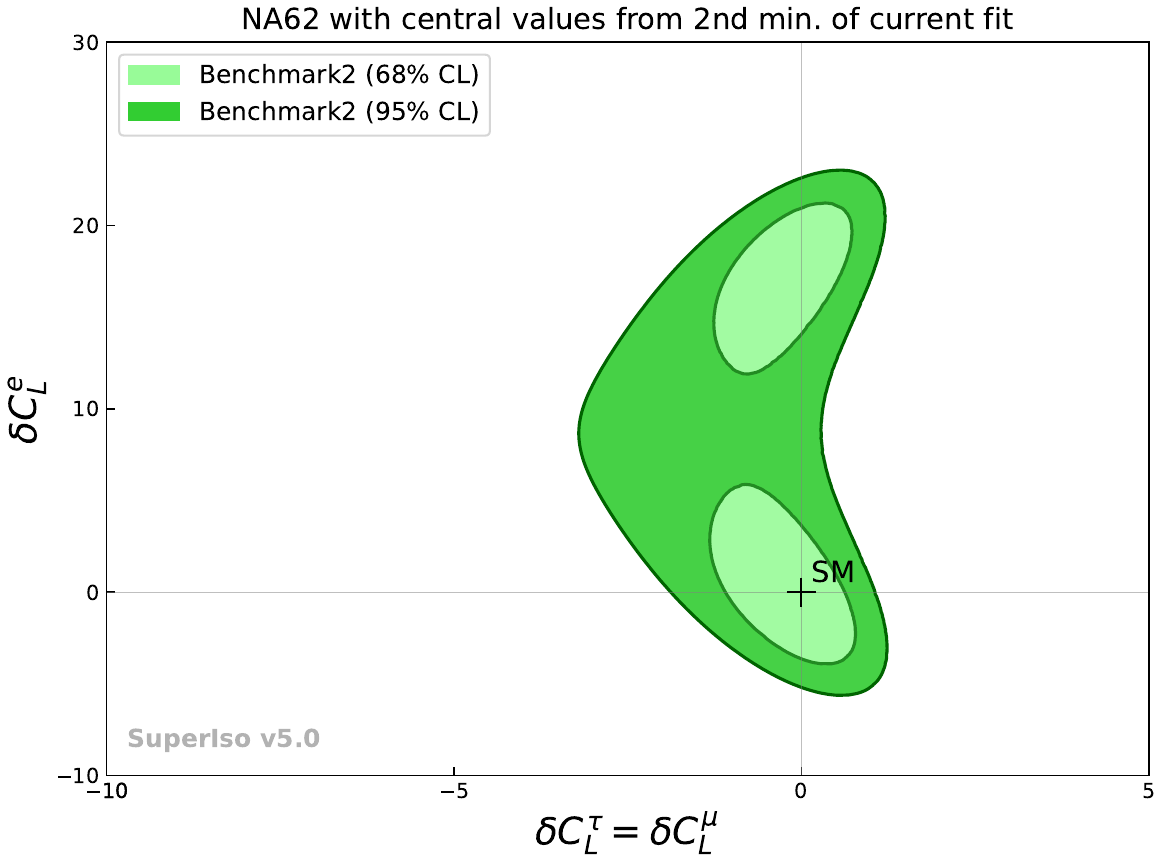}
\includegraphics[width=0.45\textwidth]{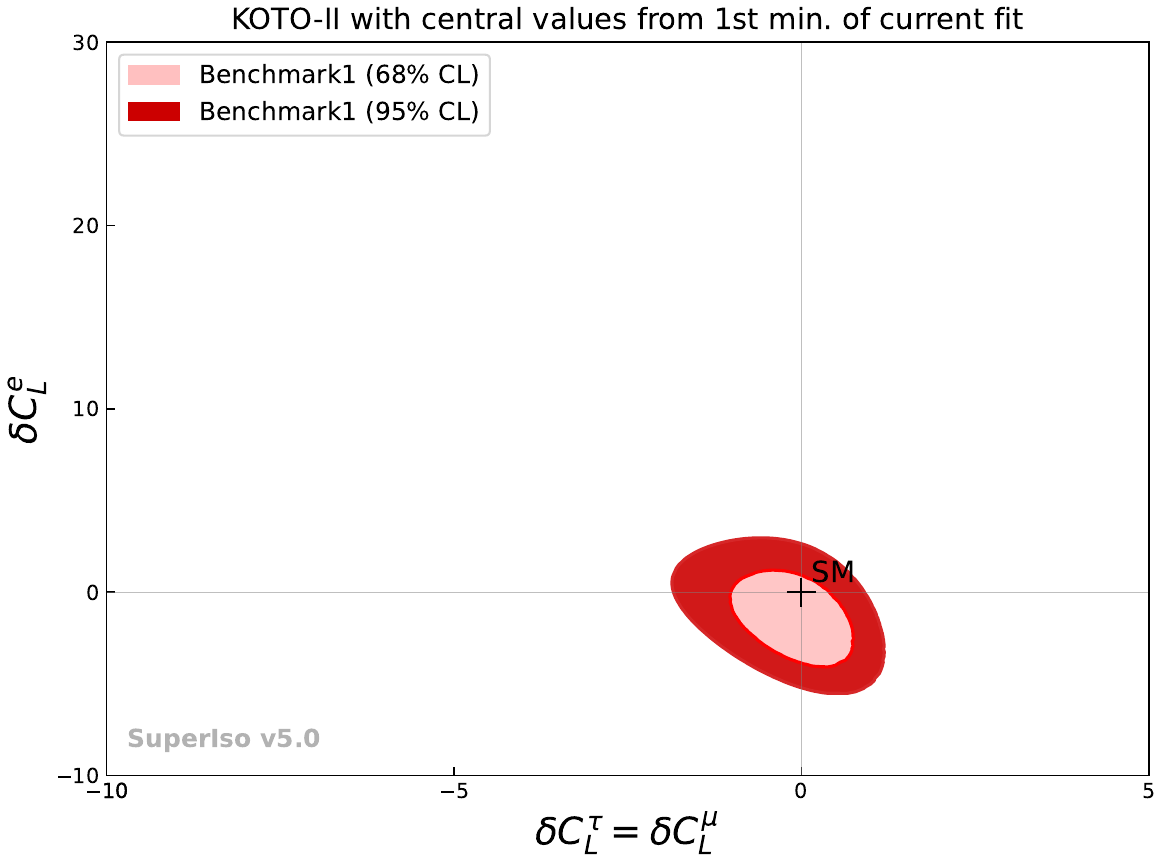}
\includegraphics[width=0.45\textwidth]{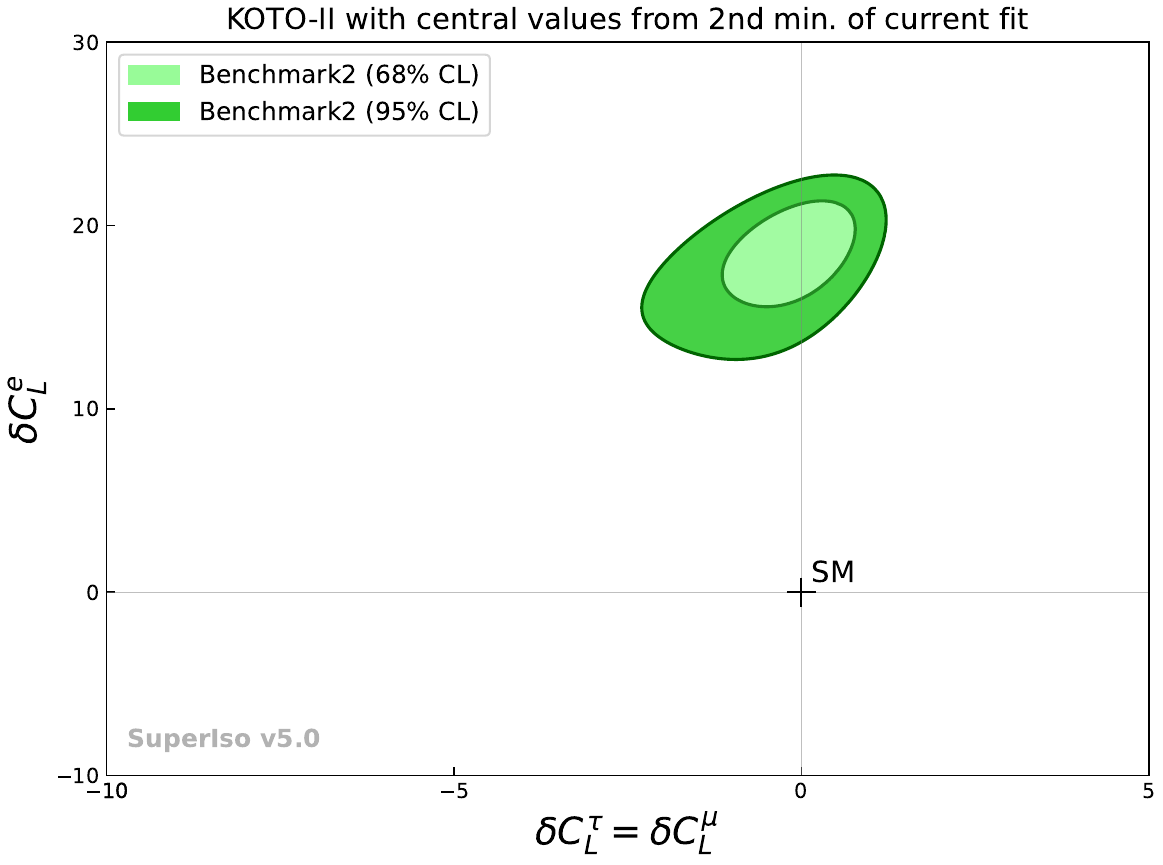}
\includegraphics[width=0.45\textwidth]{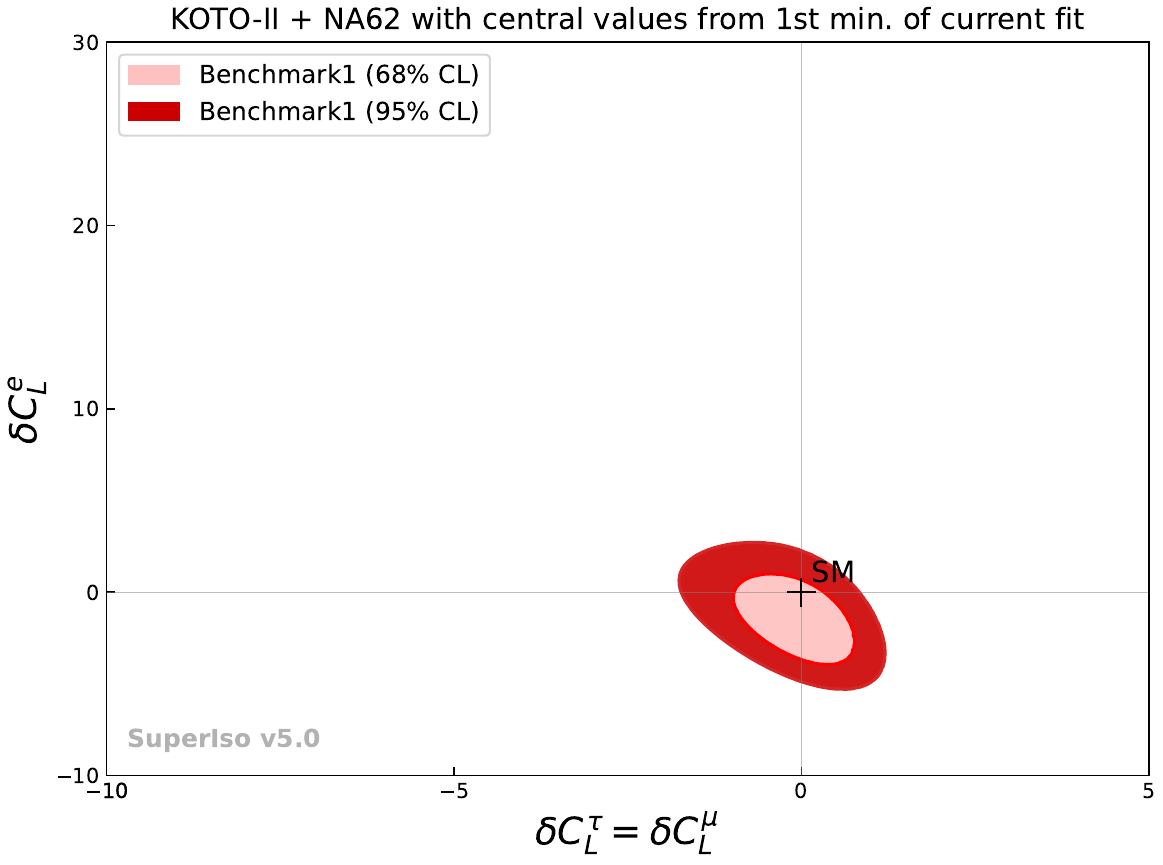}
\includegraphics[width=0.45\textwidth]{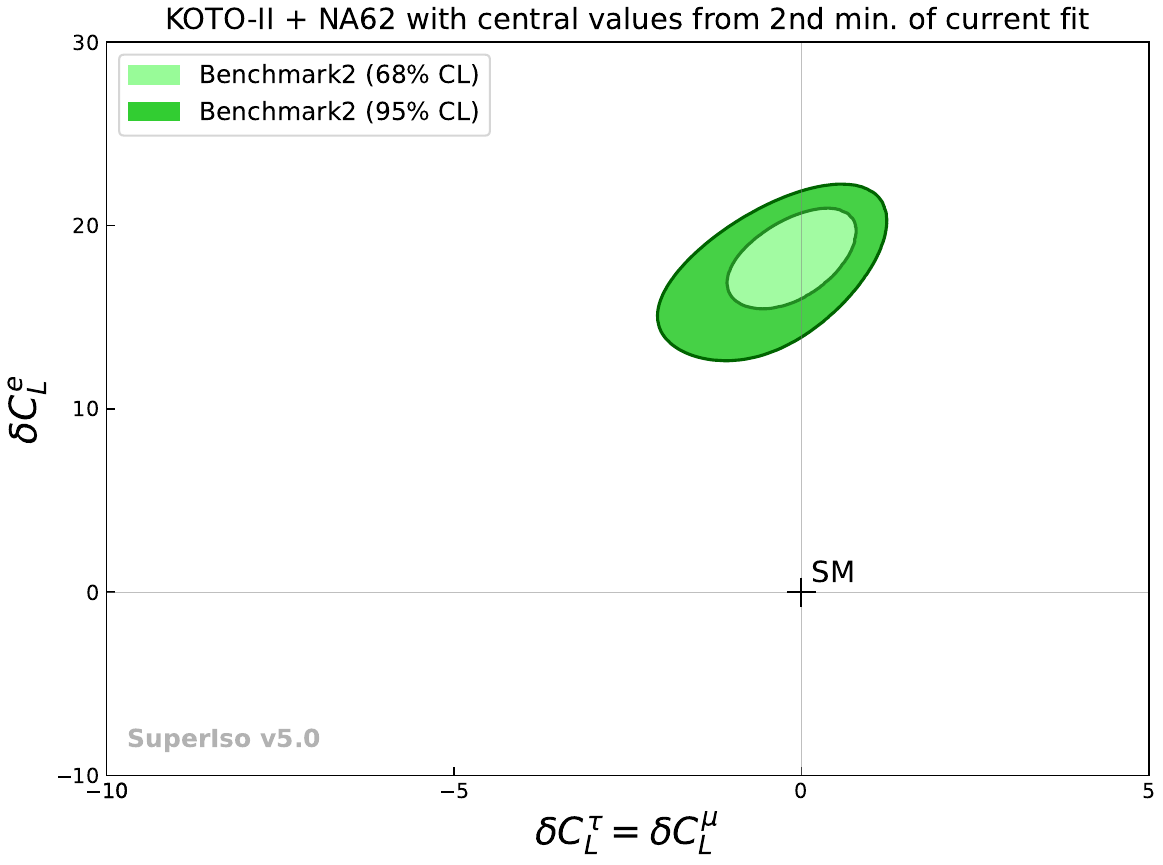}
\vspace{-0.2cm}
\caption{\small Projected constraints on the new physics parameter space. The top, middle, and bottom rows show projections from NA62 only, KOTO-II only, and the combined NA62+KOTO-II scenario, respectively. The left and right panels correspond to Benchmark 1 and Benchmark 2, respectively. The contours correspond to the $1\sigma$ and $2\sigma$ confidence levels.
\label{fig:Projections_NPerr}}
\end{center}
\end{figure}

The resulting projected constraints are shown in Fig.~\ref{fig:Projections_NPerr}. The left column corresponds to Benchmark~1 and the right column to Benchmark~2. The top row shows that the projected full NA62 dataset alone does not resolve the degeneracy between the two benchmarks. The reduced size of the $95\%$ CL regions is driven primarily by the improved precision of the LFUV observables in $K^+\to\pi^+\ell^+\ell^-$. Nevertheless, the two regions remain compatible at $95\%$ CL for both benchmarks.

The middle row shows the impact of KOTO-II alone, with the NA62 observables kept at their current measurement. The two allowed regions become separated beyond $2\sigma$ level, demonstrating the substantially greater discriminating power of the future neutral kaon measurements. In particular, the separation of the lower, lepton flavour conserving region from the upper LFUV region demonstrates the ability of KOTO-II to resolve the degeneracy left by $K^+\to\pi^+\nu\bar\nu$. 

The lower row shows the combined NA62 and KOTO-II projections. 
While the allowed regions are further reduced, the discrimination between the two scenarios remains largely driven by the KOTO-II measurements. This highlights the complementarity of the two programmes while demonstrating KOTO-II's particularly strong sensitivity to the lepton flavour structure of the new physics parameter space.

\clearpage
\section {Conclusions}
\label{sec:conclusions}

\begin{figure}[t!]
\begin{center}
\includegraphics[width=0.7\textwidth]{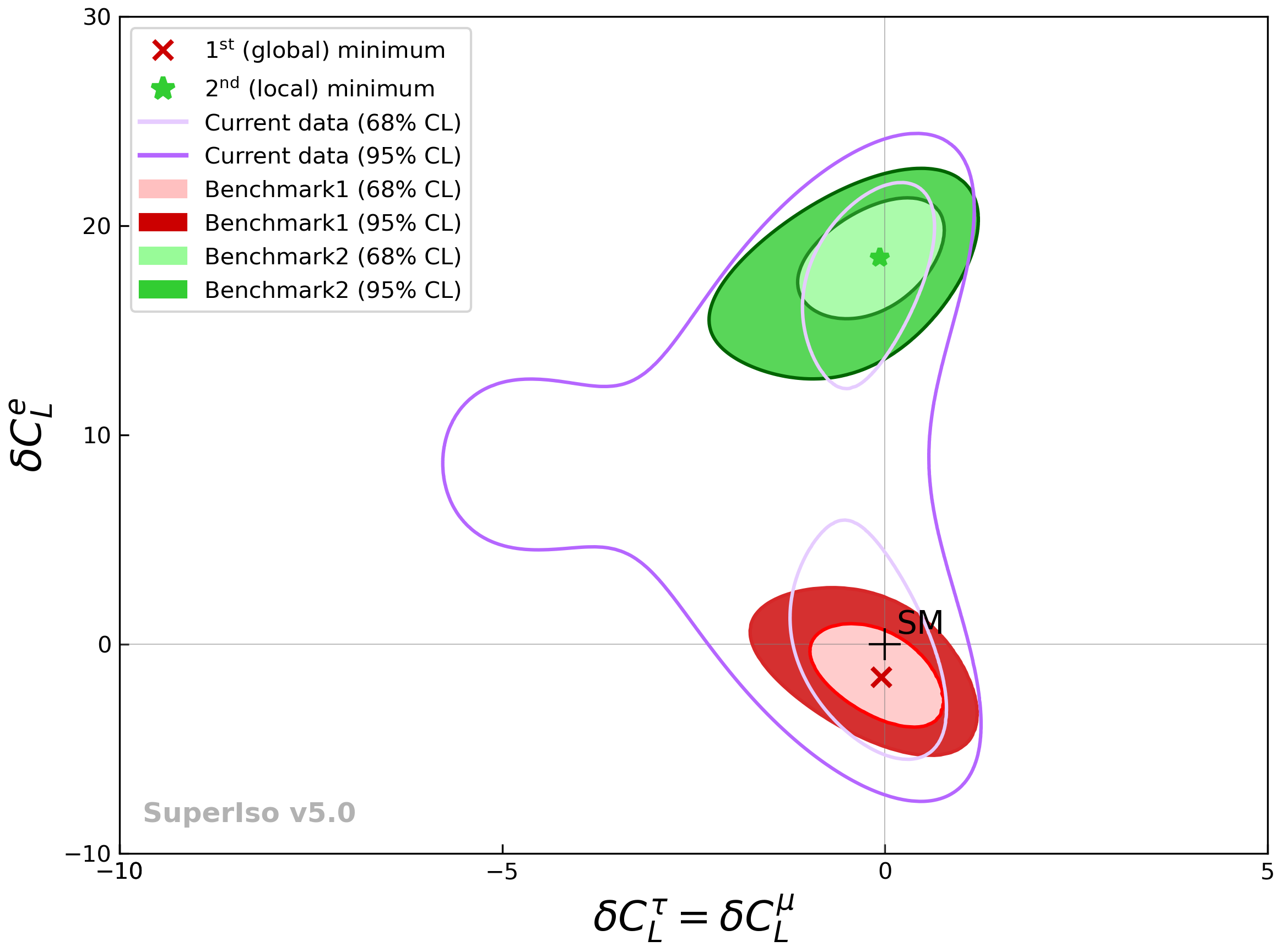}
\caption{\small Comparison of global fits to current data as well as projected fits for Benchmark~1 and~2.
\label{fig:current_projections}}
\end{center}
\end{figure}

The analysis presented in this paper is summarised in Fig.~\ref{fig:current_projections}, which illustrates the evolution from the current constraints to the sensitivity that can be achieved with the full dataset of NA62 and a future KOTO-II programme. 
In particular, the updated experimental and theoretical inputs place the current best-fit point close to the Standard Model, while a second local minimum remains and provides a phenomenologically distinct possibility for future measurements to explore. We therefore consider two benchmark scenarios corresponding to the global minimum and the second local minimum, providing a simple way to assess the ability of future experiments to distinguish between regions of parameter space that are compatible with present data.
A key highlight is the potential of KOTO-II to reveal lepton flavour universality violating effects through complementary measurements of $K_L$ decays into neutrinos and charged leptons,  while the combination with NA62 provides further sensitivity to the underlying new physics parameter space. 
If a deviation from the Standard Model were to emerge, the combination of charged- and neutral-kaon observables would provide an important tool for determining whether it originates from a lepton flavour universal contribution or from a lepton flavour dependent interaction.
This work drives home the point of pursuing kaon physics in future factories, where increased precision could possibly turn the spotlight on the parameter space where new physics lies hidden. Our results underscore the importance of pursuing rare kaon physics at future facilities, where increased experimental precision could provide a powerful window onto new physics beyond the Standard Model.

\section*{Acknowledgements}
We thank C.~Lazzeroni, H.~Nanjo and D.~Martinez Santos for useful discussions. 
This research is funded in part by the National Research Agency (ANR) under project no. ANR-21-CE31-0002-01. 

\bibliographystyle{JHEP} 
\bibliography{biblio}

\clearpage

\end{document}